\documentclass[11pt]{article}
\pdfoutput=1
\usepackage{fullpage}
\usepackage{cite}
\usepackage{graphicx}
\usepackage{amsmath}
\usepackage{amssymb}
\usepackage{subcaption}
\title{}
\date{}

\def\beq{\begin{equation}}
\def\eeq{\end{equation}}
\allowdisplaybreaks
\begin{document}
\bibliographystyle{utphys}
\newcommand{\msbar}{\ensuremath{\overline{\text{MS}}}}
\newcommand{\DIS}{\ensuremath{\text{DIS}}}
\newcommand{\abar}{\ensuremath{\bar{\alpha}_S}}
\newcommand{\bb}{\ensuremath{\bar{\beta}_0}}
\newcommand{\rc}{\ensuremath{r_{\text{cut}}}}
\newcommand{\Nd}{\ensuremath{N_{\text{d.o.f.}}}}
\setlength{\parindent}{0pt}

\titlepage
\begin{flushright}
QMUL-PH-18-26\\
\end{flushright}

\vspace*{0.5cm}

\begin{center}
{\bf \Large Biadjoint wires}

\vspace*{1cm} 
\textsc{Nadia Bahjat-Abbas\footnote{n.bahjat-abbas@qmul.ac.uk},
  Ricardo Stark-Much\~{a}o\footnote{r.j.stark-muchao@qmul.ac.uk}, 
 and Chris D. White\footnote{christopher.white@qmul.ac.uk}} \\

\vspace*{0.5cm} Centre for Research in String Theory, School of
Physics and Astronomy, \\
Queen Mary University of London, 327 Mile End
Road, London E1 4NS, UK\\

\end{center}

\vspace*{0.5cm}

\begin{abstract}
Biadjoint scalar field theory has been the subject of much recent
study, due to a number of applications in field and string theory.
The catalogue of exact non-linear solutions of this theory is
relatively unexplored, despite having a role to play in extending
known relationships between gauge and gravity theories, such as the
double copy. In this paper, we present new solutions of biadjoint
scalar theory, corresponding to singular line configurations in four
spacetime dimensions, with a power-law dependence on the cylindrical
radius. For a certain choice of common gauge group (SU(2)), a family
of infinitely degenerate solutions is found, whose existence can be
traced to the global symmetry of the theory. We also present extended
solutions, in which the pure power-law divergence is partially
screened by a form factor.
\end{abstract}

\vspace*{0.5cm}

\section{Introduction}
\label{sec:intro}

To the best of our current experimental knowledge, the forces of
nature are described by (quantum) field theories, making the latter
the subject of intense ongoing scrutiny. Recently, a number of
intriguing relationships have been discovered between field theories
whose physics is very different. One such correspondence is the {\it
  double copy}~\cite{Bern:2008qj,Bern:2010ue,Bern:2010yg}, that
relates (non-abelian) gauge theories and gravity. Although originally
formulated for scattering
amplitudes~\cite{Bern:2010ue,Bern:1998ug,Green:1982sw,Bern:1997nh,Carrasco:2011mn,Carrasco:2012ca,Mafra:2012kh,Boels:2013bi,Bjerrum-Bohr:2013iza,Bern:2013yya,Bern:2013qca,Nohle:2013bfa,
  Bern:2013uka,Naculich:2013xa,Du:2014uua,Mafra:2014gja,Bern:2014sna,
  Mafra:2015mja,He:2015wgf,Bern:2015ooa,
  Mogull:2015adi,Chiodaroli:2015rdg,Bern:2017ucb,Johansson:2015oia,Oxburgh:2012zr,White:2011yy,Melville:2013qca,Luna:2016idw,Saotome:2012vy,Vera:2012ds,Johansson:2013nsa,Johansson:2013aca},
it has subsequently been extended to classical
solutions~\cite{Monteiro:2014cda,Luna:2015paa,Luna:2016due,Goldberger:2016iau,Anastasiou:2014qba,Borsten:2015pla,Anastasiou:2016csv,Anastasiou:2017nsz,Cardoso:2016ngt,Borsten:2017jpt,Anastasiou:2017taf,Anastasiou:2018rdx,LopesCardoso:2018xes,Goldberger:2017frp,Goldberger:2017vcg,Goldberger:2017ogt,Luna:2016hge,Luna:2017dtq,Shen:2018ebu,Levi:2018nxp,Plefka:2018dpa,Cheung:2018wkq,Carrillo-Gonzalez:2018pjk,Monteiro:2018xev},
curved
space~\cite{Adamo:2017nia,Bahjat-Abbas:2017htu,Carrillo-Gonzalez:2017iyj}
and double field theory~\cite{Monteiro:2018xev}. At tree-level, the
double copy has a string theoretic
justification~\cite{Kawai:1985xq}. More generally, it has a natural
representation in terms of the so-called CHY
equations~\cite{Cachazo:2013hca,Cachazo:2013iea}, which themselves
emerge from ambitwistor string theory~\cite{Mason:2013sva}. \\

In all of the above contexts, an additional theory makes a crucial
appearance, containing a single scalar field $\Phi^{aa'}$ carrying two
adjoint indices associated with a pair of (in principle distinct) Lie
groups. This {\it biadjoint scalar field theory} can be described by
the Lagrangian density
\begin{equation}
{\cal L}=\frac12\partial^\mu\Phi^{aa'}\partial_\mu\Phi^{aa'}+\frac{y}{3}
f^{abc}\tilde{f}^{a'b'c'}\Phi^{aa'}\Phi^{bb'}\Phi^{cc'},
\label{Lagrangian}
\end{equation}
where $f^{abc}$ and $\tilde{f}^{a'b'c'}$ are the structure constants
associated with the Lie groups, and we adopt the summation convention
for repeated indices. The above cubic Lagrangian leads to the
quadratic field equation
\begin{equation}
\partial_{\mu}\partial^{\mu}\Phi^{aa'}-yf^{abc}\tilde{f}^{a'b'c'}\Phi^{bb'}\Phi^{cc'}=0.
\label{EOM}
\end{equation}
Although the biadjoint theory is not a physical theory by itself,
mounting evidence suggests that its dynamical information is
inherited, at least in part, by gauge and gravity theories. For
example, amplitudes in biadjoint scalar theory are related to those in
(non-abelian) gauge theory by a process known as the {\it zeroth
  copy}. A similar procedure holds for classical solutions, in those
cases in which the single copy between gravity and gauge theory is
also known (see the above references for further details). An
ever-increasing web of theories related by similar correspondences is
currently being established, where a recent summary can be found in
figure 1 of ref.~\cite{Carrillo-Gonzalez:2018pjk}.\\

The above correspondences involve perturbative solutions of the
respective theories, and / or those with only positive powers of the
coupling constant. It is natural to ponder whether or not the
relationships can be extended to the fully nonperturbative
regime. There are a number of ways of approaching this issue. Firstly,
one may consider how symmetries match up on both sides of the double
copy (see e.g. the recent ref.~\cite{Berman:2018hwd} for a
discussion). Secondly, one may catalogue fully non-linear solutions of
various field theories, before trying to match them up in some
way. How to do the latter is unclear, as all previous examples of how
to perform the double copy involve solutions of the linearised
biadjoint equation. However, the elucidation of new nonlinear
solutions is certainly achievable, and a necessary step in probing
nonperturbative aspects of the double copy any further. A wide
literature already exists on nonlinear solutions in gauge and gravity
theories (see
e.g. refs.~\cite{Weinberg:2012pjx,Manton:2004tk,Belinski:2001ph} for
reviews). Much less is known about exact solutions of the biadjoint
scalar theory of eq.~(\ref{EOM}).\\

Some first non-linear solutions of biadjoint theory were presented in
ref.~\cite{White:2016jzc}. They included a spherically symmetric
monopole-like object in the case in which both Lie groups are the
same, where the field $\Phi^{aa'}$ has a power-law behaviour,
diverging at the origin. An additional (and more general) solution was
found when the common gauge group is SU(2), although again possessing
spherical symmetry. Extended solutions were found in
ref.~\cite{DeSmet:2017rve}, where the power-law behaviour of the
field was dressed by a non-trivial form factor, which partially
screens (but does not remove~\footnote{Finite energy solutions are
  forbidden in scalar field theories, by Derrick's
  theorem~\cite{Derrick:1964ww}.})  the singular behaviour at the
origin. \\

The aim of this paper is to go beyond spherically symmetric solutions
of the biadjoint field equation. More specifically, we will consider
cylindrically symmetric solutions, depending only on the cylindrical
polar radius $\rho$. A number of non-trivial results will be
presented. First, we will find a power-law solution for the case in
which both Lie groups are the same. This mirrors the monopole-like
solution found in ref.~\cite{White:2016jzc}, and can be interpreted as
a wiry object localised on the $z$-axis. We will look for a more
general solution when the common Lie group is SU(2), finding a
one-parameter family of solutions, again as in
ref.~\cite{White:2016jzc} for the spherically symmetric case. Unlike
the latter, however, we will see that the family of SU(2) solutions is
degenerate in energy, which can be traced to the symmetry of the
biadjoint theory in this case. Next, we will consider dressed
solutions, for which the power-law behaviour in $\rho$ is modified by
a form factor. We will find, as in ref.~\cite{DeSmet:2017rve}, that
such form factors can indeed be obtained, and have the effect of
screening the divergent behaviour of the wire. Our results will be
important for future studies of the nonperturbative double copy, as
well as being of interest in their own right, given the multiple
contexts in which the biadjoint theory arises.\\

The structure of the paper is as follows. In section~\ref{sec:vortex},
we derive power-law line solutions for a general common gauge group
$G$, and for the case in which this group is SU(2). In
section~\ref{sec:dressed}, we consider dressed solutions. We discuss
our results and conclude in section~\ref{sec:conclude}.

\section{Power-law wire solutions}
\label{sec:vortex}

\subsection{Solution for generic common gauge group}

Let us first consider a common gauge group $G$, such that
$f^{abc}=\tilde{f}^{abc}$ in eqs.~(\ref{Lagrangian},
\ref{EOM}). Adopting cylindrical polar coordinates $(\rho,z,\phi)$, we
may look for a static cylindrically symmetric solution by making the
ansatz
\begin{equation}
\Phi^{aa'}=\delta^{aa'} f(\rho),
\label{ansatz1}
\end{equation}
where the structure constants are normalised according to
\begin{equation}
f^{abc}f^{a'bc}=T_A\delta^{aa'}.
\label{TAdef}
\end{equation}
One then finds
\begin{equation}
\frac{1}{\rho}\frac{\partial}{\partial\rho}\left(\rho
\frac{\partial f}{\partial \rho}\right)+y T_Af^2(\rho)=0, 
\label{vortexsol2}
\end{equation}
which has the non-trivial power-law solution
\begin{equation}
\Phi^{aa'}=-\frac{4\delta^{aa'}}{y T_A \rho^2}.
\label{Phivortex}
\end{equation}
As for the spherically symmetric monopole-like object found in
ref.~\cite{White:2016jzc}, this has an inverse power of the coupling
constant $y$, and thus is nonperturbative (i.e. it vanishes at weak
coupling). It goes like the inverse square of the cylindrical radius
$\rho$, where this dependence can also be surmised by dimensional
analysis. The solution is singular as $\rho\rightarrow 0$,
corresponding to a line defect in the field, localised on the
$z$-axis. To further illustrate this, we may calculate the energy of
the field, for which eq.~(\ref{Lagrangian}) implies the Hamiltonian
density (see ref.~\cite{White:2016jzc} for further details)
\begin{equation}
{\cal H}=\frac{160}{3}\frac{{\cal N}}{y^2 T_A^2}\frac{1}{\rho^6},
\label{Phisol1}
\end{equation}
where ${\cal N}$ is the dimension of the common gauge group $G$. This
diverges as $\rho\rightarrow 0$ due to the singularity in the field,
but is well-behaved as $\rho\rightarrow\infty$. We can thus place a
cutoff $\rho_0$ around the wire, and calculate an energy per unit
length
\begin{equation}
\frac{E}{L}=\frac{80\pi {\cal N}}{3 y^2 T_A^2 \rho_0^4}.
\label{ELsol1}
\end{equation}

\subsection{Solutions for SU(2)$\times$ SU(2)}
\label{sec:SU(2)}

For the special case in which the common gauge group $G$ is SU(2), one
may write a more general form for the field. The structure constants
in that case are equal to the Levi-Cevita tensor
$f^{abc}=\epsilon^{abc}$, allowing the possibility of mixing spatial
and gauge indices. This was used in
refs.~\cite{White:2016jzc,DeSmet:2017rve} to find novel spherically
symmetric solutions, and a similar (but not identical) ansatz can be
used here. A notable feature in the present case -- which is shared by
similar solutions in Yang-Mills theory~\cite{Sikivie:1978sa} -- is
that the requirement of cylindrical symmetry, together with the mixing
of spatial and gauge indices, means that choosing a special direction
in space (the $z$-axis) picks out a special direction (the
3-direction) in the gauge space, suggesting the following ansatz:
\begin{equation}
\Phi^{33}=f_1(\rho),\quad 
\Phi^{ij}=f_2(\rho)\delta^{ij}+f_3(\rho)x^i x^j+f_4(\rho)\epsilon^{3ij}, \quad
\Phi^{3i}=\Phi^{i3}=0.
\label{Phi33}
\end{equation}
Here and in the following, we use indices $i,j,k\ldots\in(1,2)$, as
distinct from indices $a,b,c\in(1,2,3)$. Substituting
eq.~(\ref{Phi33}) into eq.~(\ref{EOM}), one obtains the four coupled
non-linear ordinary differential equations
\begin{align}
&\frac{1}{\rho}\frac{\partial}{\partial \rho}\left(\rho\frac{\partial
  \bar{f}_2} {\partial \rho}\right)+2 \bar{f}_3+2 \bar{f}_1\left(\bar{f}_2
+\rho^2 \bar{f}_3\right)=0;\notag\\
&\frac{\partial^2\bar{f}_3}{\partial \rho^2}+\frac{5}{\rho}
\frac{\partial \bar{f}_3}{\partial \rho}-2\bar{f}_1\,\bar{f}_3=0;\notag\\
&\frac{1}{\rho}\frac{\partial}{\partial\rho}\left(\rho
\frac{\partial \bar{f}_4}{\partial\rho}\right)+2\bar{f}_1\,\bar{f}_4=0;\notag\\
&\frac{1}{\rho}\frac{\partial}{\partial\rho}\left(\rho\frac{\partial \bar{f}_1}
{\partial\rho}\right)+2\left(\bar{f}_2^2+\bar{f}_4^2
+\rho^2\bar{f}_2\,\bar{f}_3\right)=0,
\label{feqs}
\end{align}
where, following ref.~\cite{White:2016jzc}, we have defined the
convenient combinations
\begin{equation}
f_i(\rho)=\frac{\bar{f}_i(\rho)}{y}.
\label{fbardef}
\end{equation}
We may find a power-law solution to eqs.~(\ref{feqs}) by writing
\begin{equation}
\bar{f}_i=k_i \rho^{\alpha_i},
\label{kalpha}
\end{equation}
substitution of which straightforwardly yields
\begin{equation}
\alpha_1=\alpha_2=\alpha_4=-2,\quad \alpha_3=-4,
\label{alphavals}
\end{equation}
as well as the nonlinear simultaneous equations
\begin{align}
& 2k_2+k_3+k_1(k_2+k_3)=k_4(2+k_1)=
2k_1+k_2^2+k_4^2+k_2\,k_3=k_1\,k_3=0.
\label{keqs}
\end{align}
Setting $k_2=-2k$, the general solution of these equations implies 
\begin{equation}
\Phi^{33}=-\frac{2}{y\rho^2},\quad
\Phi^{ij}=-\frac{2}{y\rho^2}\left[k\delta^{ij}\mp \sqrt{1-k^2}
\epsilon^{3ij}\right].
\label{Phi33sol}
\end{equation}
There is thus a continuously infinite family of solutions. 
Note that the previous solution of eq.~(\ref{Phivortex}) emerges as
the special case $k=1$. As for that case, one may calculate the energy
associated with eq.~(\ref{Phi33sol}), subject to a cutoff being
applied for low cylindrical radius. An explicit calculation reveals an
energy per unit length
\begin{equation}
\frac{E}{L}=\frac{20\pi}{y^2\rho_0^4},
\label{ELsol2}
\end{equation}
regardless of the value of $k$. Hence, the family of solutions
obtained in eq.~(\ref{Phi33sol}) is degenerate. Furthermore,
eq.~(\ref{ELsol2}) agrees with eq.~(\ref{ELsol1}) for the specific
SU(2) values ${\cal N}=3$, $T_A=2$, as it should given the degeneracy,
and the fact that the previous solution emerges from the present one
as a special case. \\

One may understand the degeneracy of the family of solutions of
eq.~(\ref{Phi33sol}) as follows. First, we may write $k=\cos\theta$,
and introduce a matrix ${\bf \Phi}$, whose components are
$\Phi^{aa'}$:
\begin{equation}
{\bf \Phi}=-\frac{2}{y\rho^2}\left(\begin{array}{ccc}
\cos\theta & \mp \sin\theta & 0 \\
\pm \sin\theta & \cos\theta & 0\\
0 & 0 & 1\end{array}\right).
\label{Phimat}
\end{equation}
This is a rotation about the $3$-axis in gauge space by angle
$\theta$.  Next, we may write the Lagrangian of
eq.~(\ref{Lagrangian}), for the special case of SU(2), as
\begin{align}
{\cal L}&=\frac12\partial^\mu\Phi^{aa'}\partial_\mu\Phi^{aa'}+\frac{y}{3}
\epsilon^{abc}
\epsilon^{a'b'c'}\Phi^{aa'}\Phi^{bb'}\Phi^{cc'}
=\frac12 {\rm Tr}\left[(\partial^\mu{\bf \Phi})^{\rm T}
(\partial_\mu{\bf \Phi})\right]+2\,y\,{\rm det}[{\bf \Phi}].
\label{LSU2}
\end{align}
This is manifestly invariant under the transformation ${\bf
  \Phi}\rightarrow {\bf R}^{\rm T}_1{\bf \Phi}{\bf R}_2$, for
arbitrary rotation matrices ${\bf R}_i$. Thus, one may start with the
solution of eq.~(\ref{Phivortex}) for SU(2), which in the present
notation reads
\begin{equation}
{\bf \Phi}=-\frac{2}{y\rho^2}\left(
\begin{array}{ccc}1 & 0 & 0 \\ 0 & 1 & 0 \\ 
0 & 0 & 1 \end{array}\right),
\label{Phimatsol1}
\end{equation}
before acting on it with rotations about the 3-axis to generate the
family of solutions of eq.~(\ref{Phi33sol}). The rotation will not
change the energy, given the above-mentioned symmetry, which is none
other than the global symmetry associated with the SU(2) biadjoint
theory~\footnote{A similar interpretation can be applied to the
  spherically symmetric SU(2)$\times$SU(2) solution in
  ref.~\cite{White:2016jzc}, which consists of an infinitesimal
  rotation of the analogue of eq.~(\ref{Phivortex}).}.

\section{Dressed wires}
\label{sec:dressed}

In the preceding section, we have found pure power-law wire solutions
of the biadjoint theory. This is not the whole story, however. In
gauge theories, there is a cornucopia of solutions in which a pure
power-law divergence can be dressed with a non-trivial form factor,
where the latter can be interpreted as providing some internal
structure (see
e.g. refs.~\cite{Weinberg:2012pjx,Manton:2004tk}). Furthermore, it can
have the effect of screening the divergent behaviour near singular
regions of the field. Reference~\cite{DeSmet:2017rve} found that such
solutions also exist in the biadjoint theory, namely that the
spherically symmetric monopole solution can be dressed with a
screening function. Given that our aim is to catalogue - both
qualitatively and quantitatively - the solutions that are possible in
biadjoint theory, it is both interesting and useful to check whether
or not such dressed behaviour is possible also for the wire
solution. \\

To this end, let us reconsider the ansatz of eq.~(\ref{ansatz1}), and
define~\footnote{Strictly speaking, the left-hand side of
  eq.~(\ref{Jdef}) should be a function of a dimensionless ratio
  $\rho/\rho_0$ for some length scale $\rho_0$, but we may take the
  latter to be unity without loss of generality in what follows.}
\begin{equation}
J(\rho)=2+\rho^2 f(\rho).
\label{Jdef}
\end{equation}
Dimensional analysis fixes $f(\rho)\sim \rho^{-2}$, so that $J(\rho)$
must be finite for all $\rho$. Substituting eq.~(\ref{Jdef}) into
eq.~(\ref{vortexsol2}) yields~\footnote{Varying the constant term on
  the right-hand side of eq.~(\ref{Jdef}) would lead to an additional
  term linear in $K$, which it is convenient to remove by the above
  choice.}
\begin{equation}
\rho^2\frac{d^2 J(\rho)}{d \rho^2} -3\rho\frac{d
  J(\rho)}{d\rho}+J(\rho)^2-4=0.
\label{Jeq1}
\end{equation}
Upon transforming to $\xi$ via
\begin{equation}
\rho=e^{-\xi},
\label{xidef}
\end{equation}
eq.~(\ref{Jeq1}) becomes an ODE with constant coefficients:
\begin{equation}
\frac{d^2 J(\xi)}{d\xi^2}+4\frac{dJ(\xi)}{d\xi}+J^2(\xi)-4=0.
\label{Jeq2}
\end{equation}
This can be related to an Abel equation of the second kind, albeit one
with no analytic solution (see also ref.~\cite{DeSmet:2017rve}). We
may instead look for a numerical solution, as follows. One may write
eq.~(\ref{Jeq2}) as a set of coupled first order equations by defining
\begin{equation}
\left(\frac{d J}{d\xi},\frac{d\psi}{d\xi}\right)=
\left(\psi,4-4\psi-J^2\right).
\label{Jvec}
\end{equation}
Next, one may plot the integral curves of this vector field in the
$(J,\psi)$ plane, where any bounded curves correspond to solutions for
$J(\xi)$ that remain finite for all $\xi$ (and thus $\rho$). We show
this vector field in figure~\ref{fig:vectorfield}, whose examination
yields the following solutions:
\begin{figure}
\begin{center}
\scalebox{0.4}{\includegraphics{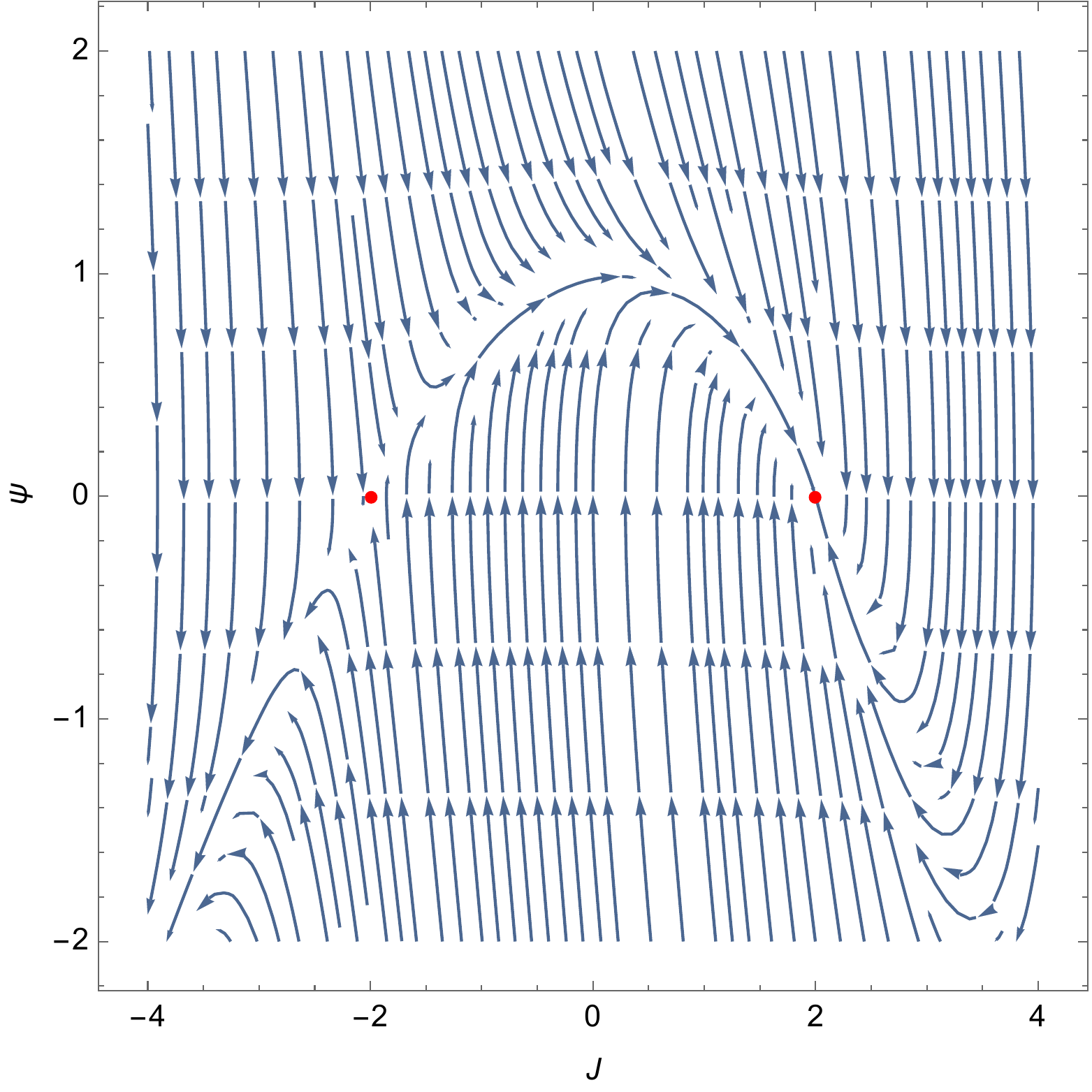}}
\caption{Integral curves of the vector field of
  eq.~(\ref{Jvec}). Bounded curves (or fixed points) correspond to
  solutions for $J(\xi)$ that are finite for all $\xi$.}
\label{fig:vectorfield}
\end{center}
\end{figure}
\begin{enumerate}
\item $J(\xi)=-2$: this corresponds to the solution of
  eq.~(\ref{Phivortex}).
\item $J=2$: this yields the trivial solution $\Phi^{aa'}=0$.
\item $J(\xi)\rightarrow\pm 2$ as $\xi\rightarrow \pm \infty$
  respectively. This is a non-trivial form factor, corresponding to
  the curve that flows from $(-2,0)$ to $(2,0)$ in
  figure~\ref{fig:vectorfield}.
\end{enumerate}
It is worth remarking that the structure of
figure~\ref{fig:vectorfield} is qualitatively similar to that of the
spherically symmetric case of ref.~\cite{DeSmet:2017rve}, although the
non-trivial form factor (in case 3) $J(\xi)$ is different. We can
solve for $J(\xi)$ in the asymptotic limits $\xi\rightarrow\pm\infty$
by writing
\begin{equation}
J(\xi)=\pm 2+\chi_\pm(\xi),\quad \xi\rightarrow\pm\infty.
\label{Jlims}
\end{equation}
In each limit, we may neglect terms in $\chi_\pm^2$, obtaining
approximate solutions (for finite $J(\xi)$)
\begin{equation}
J(\xi)\simeq \begin{cases} 
-2+c_1 e^{\left(2\sqrt{2}-2\right)\xi},
&\quad \xi\rightarrow-\infty\\
+2+c_2e^{-2\xi}+c_3\xi e^{-2\xi},&\quad \xi\rightarrow+\infty
\end{cases}
\label{Klims}
\end{equation}
or in terms of $\rho$:
\begin{equation}
J(\rho)\simeq \begin{cases} 
-2+c_1 \rho^{2-2\sqrt{2}},
&\quad \rho\rightarrow \infty\\
+2+(c_2-c_3\log\rho)\rho^{2},&\quad \rho\rightarrow0.
\end{cases}
\label{Klims2}
\end{equation}
We can solve for the complete form of $J(\xi)$ numerically, upon
choosing $c_1=1$~\footnote{A different choice of $c_1$ amounts to
  shifting $\xi$ by a constant, or rescaling $\rho$, neither of which
  affect the qualitative shape of figure~\ref{fig:Jplot}.}. We plot
this numerical solution in figure~\ref{fig:Jplot}.
\begin{figure}
\begin{center}
(a)\scalebox{0.4}{\includegraphics{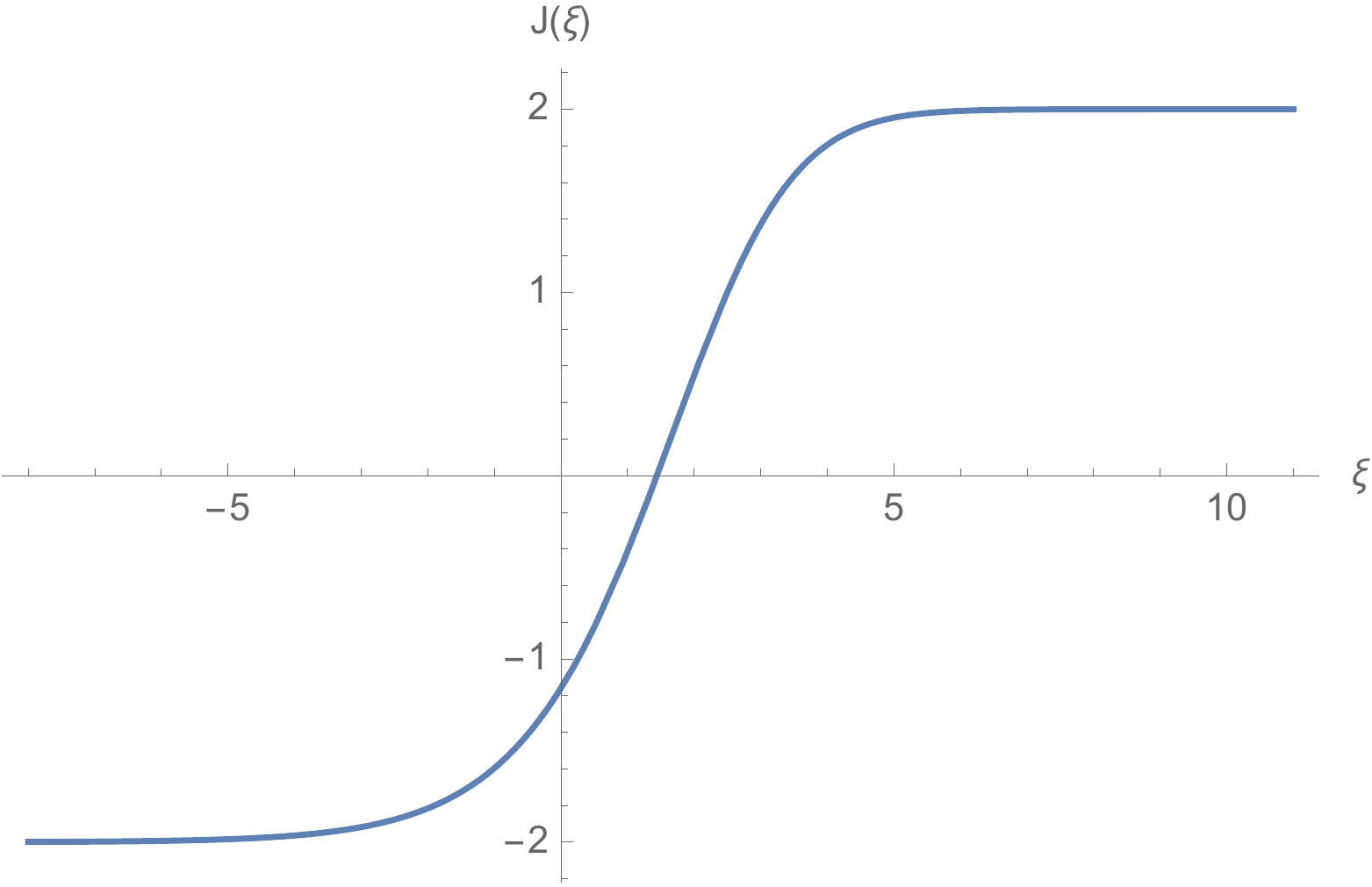}}\hspace{2cm}
(b)\scalebox{0.4}{\includegraphics{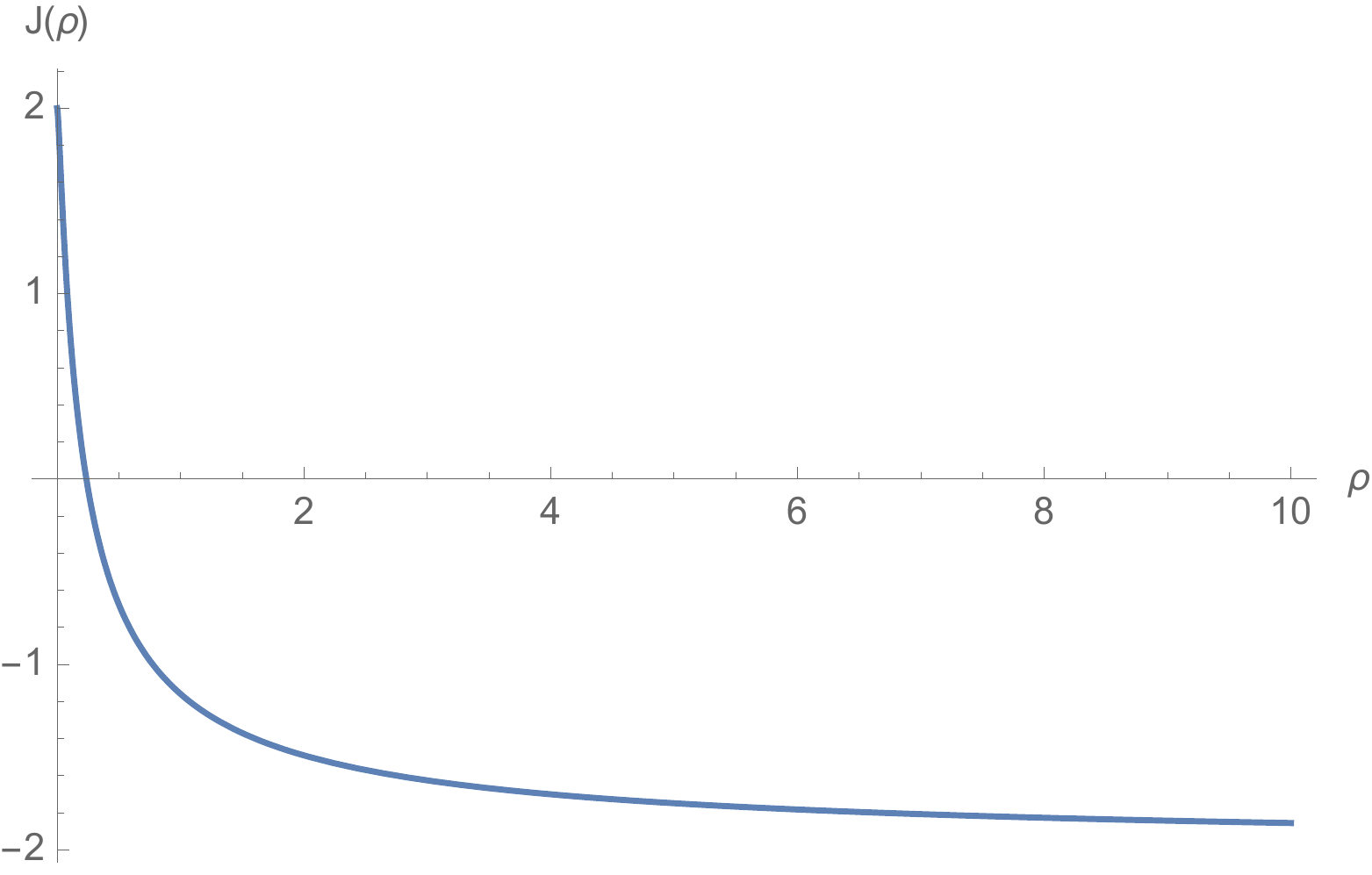}}
\caption{(a) Numerical solution of $J(\xi)$ from eq.~(\ref{Jdef}),
  with boundary conditions as described in the text; (b) Behaviour of
  $J$ as a function of the cylindrical radius $\rho$.}
\label{fig:Jplot}
\end{center}
\end{figure}
The boundary condition $J(\rho)\rightarrow 2$ as $\rho\rightarrow 0$
means that the divergence of the wire solution near the $z$-axis is
partially screened. Indeed, we find an energy per unit length of
\begin{equation}
\frac{E}{L}=\frac{\pi c_3^2{\cal N}}{T_A^2 y^2}\frac{1}{\rho_0}
+{\cal O}(\rho^0),
\label{ELdressed}
\end{equation}
which is less singular than the undressed result of
eq.~(\ref{ELsol1}). In order to visualise the dressed wire, we show
in figure~\ref{fig:density} the combination $\rho^2 \bar{f}(\rho)$ in the
$(x,y)$ plane i.e. the profile function of eq.~(\ref{ansatz1}), with
the singular factor $\rho^2$ removed. This corresponds to looking down
on the wire, which is situated at the origin. 
\begin{figure}
\begin{center}
\scalebox{0.4}{\includegraphics{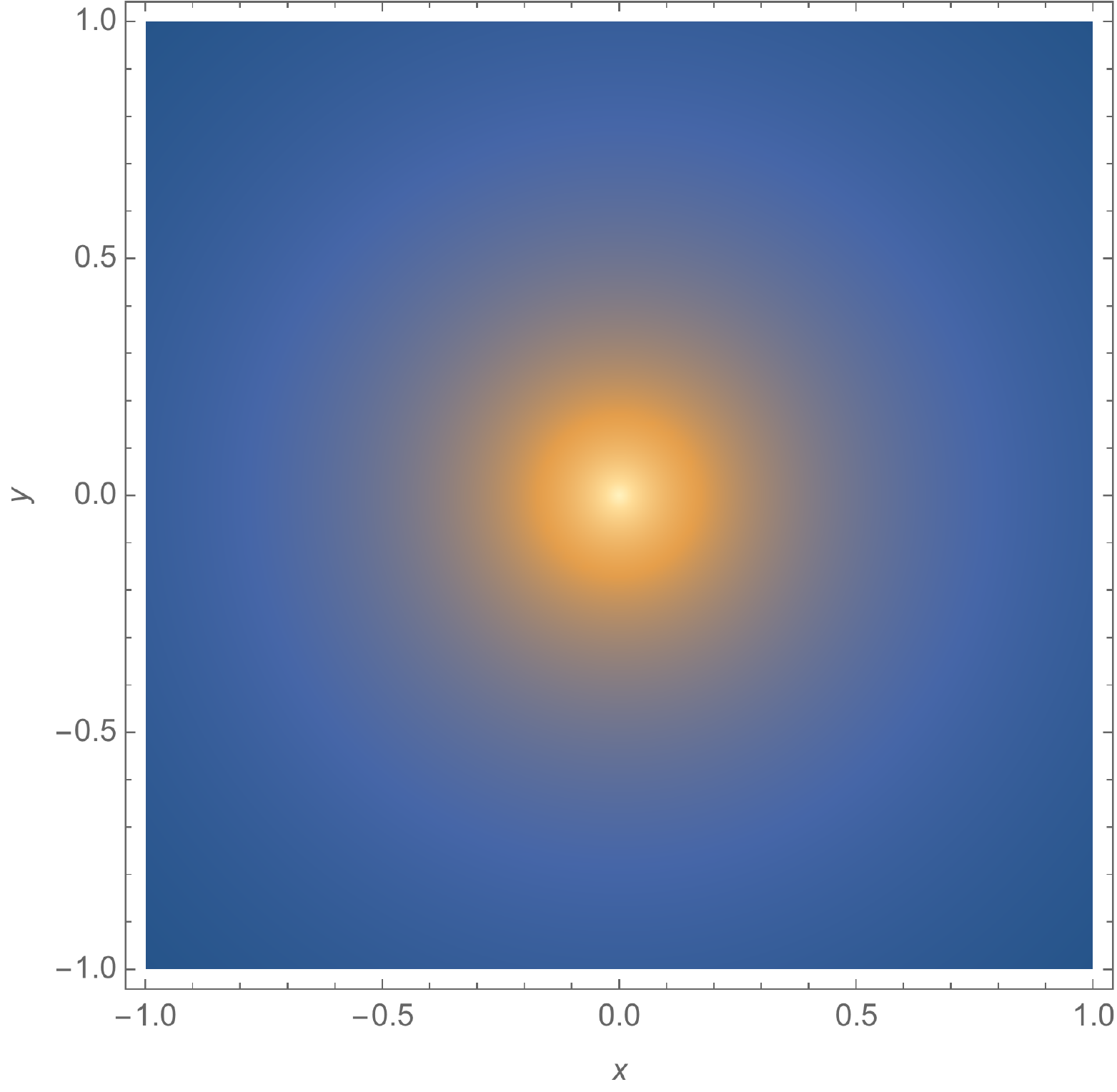}}
\caption{The profile function $\rho^2 f(\rho)$ in the $(x,y)$ plane
  for a dressed wire solution situated on the $z$-axis.}
\label{fig:density}
\end{center}
\end{figure}
Finally, we note that it would also be possible to generate dressed
solutions in the SU(2)$\times$SU(2) theory of section~\ref{sec:SU(2)},
by rotating the solution obtained here, as in eq.~(\ref{Phimat}).

\section{Conclusion}
\label{sec:conclude}

In this paper, we have obtained new solutions of the biadjoint scalar
field theory, which occurs in a number of contexts, including the
double copy relationship between gauge theories and gravity. Although
much is already known about the perturbative sector of biadjoint
theory, much less is known about its nonperturbative properties,
making any results of interest in their own right. Furthermore,
solutions such as that found in this study open up the possibility to
try to extend the double copy to a fully nonperturbative regime.\\

We have focused specifically on cylindrically symmetric solutions,
finding a power-law solution corresponding to a wire-like object
localised on the $z$-axis, for the case in which both Lie groups in
the biadjoint theory are the same. For the case in which this common
group is SU(2), an infinite family of solutions is possible. Such
solutions are degenerate, an effect which can be traced to the global
symmetry of the biadjoint theory for this choice of gauge group. \\

As well as pure power-law solutions, we also constructed dressed
wires, in which a form factor partially screens the divergent
behaviour of the field near the core. This is further evidence of a
potentially rich spectrum of nonperturbative solutions in biadjoint
theory, whose properties mimic those found in gauge theories. Whether
or not the relationship with gauge theory can be made precise is the
subject of ongoing research, and we hope that the results of this
investigation will prove useful in this regard.

\section*{Acknowledgments}

We thank David Berman, Ricardo Monteiro and Costis Papageorgakis for
useful comments and discussions. NBA and RSM are supported by PhD
studentships from the United Kingdom Science and Technology Facilities
Council (STFC) and the Royal Society respectively. This research was
also funded by STFC grant ST/P000754/1.


\bibliography{refs.bib}
\end{document}